\documentclass[conference,transmag]{IEEEtran}
\ifCLASSINFOpdf
\else
\fi

\usepackage{graphicx}
\usepackage[
top    = 0.7in,
bottom = 0.95in,
left   = 0.57in,
right  = 0.57in]{geometry}\graphicspath{{images/}}
\usepackage{svg}
\usepackage{amsmath}
\usepackage{fancyhdr}
\usepackage{caption}
\usepackage[linesnumbered,ruled]{algorithm2e}

\usepackage{amsmath}
\usepackage{flushend}
\usepackage{textcomp}
\usepackage{algorithmicx}

\usepackage{subfiles}
\usepackage{tabularx}
\usepackage{caption}
\usepackage{subcaption}

\makeatletter
\def\BState{\State\hskip-\ALG@thistlm}
\makeatother

\usepackage[
    backend=biber, 
    natbib=true,
    style=numeric,
    firstinits,
    sorting=none,
    maxnames=100
]{biblatex}

\renewbibmacro{in:}{\ifentrytype{article}{}{\printtext{\bibstring{in}\intitlepunct}}}

\makeatletter
\newcommand*\titleheader[1]{\gdef\@titleheader{#1}}
\AtBeginDocument{%
  \let\st@red@title\@title
  \def\@title{%
    \bgroup\normalfont\large\centering\@titleheader\par\egroup
    \vskip0.1em\st@red@title}
}

\begin{document}

\title{Towards Immutability: A Secure and Efficient Auditing Framework for Cloud Supporting Data Integrity and File Version Control}



\author{\IEEEauthorblockN{Faisal Haque Bappy$^{1}$, Saklain Zaman$^{1}$, Tariqul Islam$^{1}$, Redwan Ahmed Rizvee$^{2}$, Joon S. Park$^{1}$, and Kamrul Hasan$^{3}$}
\IEEEauthorblockA{$^{1}$ School of Information Studies (iSchool), Syracuse University, Syracuse, NY, USA \\
$ ^{2}$ University of Dhaka, Bangladesh \\
$ ^{3}$ Tennessee State University, Nashville, TN, USA \\
Email: $\lbrace$\textit{\{fbappy, szaman02, mtislam, jspark@\}@syr.edu, rizveeredwan.csedu@gmail.com, mhasan1@tnstate.edu}$\rbrace$}}

	
\maketitle

\thispagestyle{fancy}
\lhead{This work has been accepted at the 2023 IEEE Global Communications Conference (GLOBECOM)}
\cfoot{}

\begin{abstract}

 \textit{Abstract---}Although wide-scale integration of cloud services with myriad applications increases quality of services (QoS) for enterprise users, verifying the existence and manipulation of stored cloud information remains an open research problem. Decentralized blockchain-based solutions are becoming more appealing for cloud auditing environments because of the immutable nature of blockchain. However, the decentralized structure of blockchain results in considerable synchronization and communication overhead, which increases maintenance costs for cloud service providers (CSP). This paper proposes a Merkle Hash Tree based architecture named \texttt{Entangled Merkle Forest} to support version control and dynamic auditing of information in centralized cloud environments. We utilized a semi-trusted third-party auditor to conduct the auditing tasks with minimal privacy-preserving file-metadata. To the best of our knowledge, we are the first to design a node sharing Merkle Forest to offer a cost-effective auditing framework for centralized cloud infrastructures while achieving the immutable feature of blockchain, mitigating the synchronization and performance challenges of the decentralized architectures. Our proposed scheme outperforms it's equivalent Blockchain-based schemes by ensuring time and storage efficiency with minimum overhead as evidenced by performance analysis.

\end{abstract}


\section{Introduction}
In the sectors of industrial manufacturing, vehicular systems, supply chains, smart homes, and medical care, for example, the integration of the cloud services is becoming more prevalent \cite{wang}. Although taking advantage of the computing and communication resources provided by cloud services improves user experience and quality of service (QoS), limited physical control over user data raises data security concerns related to unauthorized modification or rarely accessed data deletion by a centralized server. Proving information availability and checking information tampering are mandatory requirements for cloud services as CSP might act maliciously to discard rarely accessed user data to reduce cloud storage maintenance costs \cite{yu}. In order to gain the trust of users to store sensitive information in centralized cloud storage systems, cloud service providers must ensure the auditability of stored information. Once user data is uploaded and exposed to the cloud, cloud service providers are required to maintain the integrity and confidentiality of data \cite{garg}. The literature discusses various cloud service integrity verification approaches, such as employing third-party auditors to perform public auditing on behalf of the users \cite{garg}\cite{zhu}, integrating edge and fog computing \cite{wang}\cite{zhang}, deploying blockchain-based auditing frameworks in the cloud environment \cite{shaf}\cite{jiax}\cite{ped}, and so on. 

The transparent and immutable nature of blockchain-based decentralized solutions is being studied as they can certify proof of existence and detect unauthorized alterations of cloud storage information \cite{fran}. As peer-to-peer systems retain information collectively, rewarding the honest behavior of participant nodes is essential to guarantee active involvement of participants, which may increase maintenance costs in permissioned cloud auditing frameworks \cite{jiax}. The time-intensive coordination and synchronization of multiple nodes in decentralized solutions may increase the cost of auditing for cloud service providers \cite{garg}.

Although the immutability of blockchain makes it suitable for data auditing environments, blockchain-based solutions suffer from computation and performance issues \cite{deep}. If immutability is attained in centralized cloud architectures, performance issues in decentralized blockchain-based solutions can be addressed while providing secure, tamper-proof, and auditable cloud storage. In order to achieve this goal, in this paper, we propose an immutable centralized cloud architecture that can facilitate version control and dynamic auditing of user data with the assistance of a semi-trusted third-party auditor. The computationally intensive auditing tasks are delegated to semi-trusted third-party auditors (TPA), requiring only lightweight computation at the user end to verify data integrity. Third-party auditor (TPA) conducts the auditing process with user provided minimal file metadata, prohibiting TPA's access to the original file owned by users and thus, ensuring data confidentiality. 

The following are the major contributions of this work:
\begin{itemize}
\item We address the synchronization overhead of decentralized blockchain solutions in auditing environments by achieving immutability in centralized cloud architecture. 

\item  We propose a Merkle Tree based novel architecture named \textbf{Entangled Merkle Forest} with node sharing to enable version control and dynamic auditing of the information stored in the cloud.
\item The semi-trusted auditor in our proposed scheme can perform batch auditing in a secure and efficient way. Since file-metadata is first encrypted and then hashed, our scheme preserves confidentiality and integrity of sensitive user information. 

\item We have implemented a prototype of dynamic auditing framework and compared it's performance with two equivalent baseline blockchain-based auditing schemes. Performance evaluation proves that our scheme outperforms both schemes while providing security, immutability, and transparency features of Blockchain. 
\end{itemize}

The remainder of the paper is organized in the following order. The status of our work compared to the literature is presented in section II. The overview of the proposed architecture is described in section III. In section IV, we present our implementation overview and then the system evaluation is presented in section V. Finally, in section VI, we conclude the paper.

\section{Related Works}
Zhu et al. propose a short signature-based architecture for data integrity verification in cloud environments \cite{zhu}. However, this architecture depends on random masking techniques for signature generation and doesn’t support data auditing for multi-replica cloud environments. Furthermore, relying on a hash or digital signature-based approach for information auditing can be inappropriate for conducting integrity verification in cloud servers when users don’t maintain a backup copy of their data. Garg et al. introduce a Merkle Hash Tree-based public data auditing scheme that can ensure the freshness of information, support data dynamic procedures, and address the privacy and integrity of outsourced data by leveraging third-party auditors (TPA) \cite{garg}. In this work, the third-party auditors (TPA) are presumed to be trustworthy which implies that the TPA won’t be curious about users’ cloud data. In our proposed architecture, we address this issue by performing auditing with user-provided minimal metadata, and thus better preservation of information privacy is achieved by controlling the prerequisite knowledge required by the TPA to perform the auditing so that the TPA can’t trace back to the original user data. 

Tian et al. employ fog nodes as a layer between the cloud server and clients to minimize the computation overhead \cite{tian}. The responsibility of encrypting data blocks is assigned to the fog nodes which are assumed to be credible and trustworthy but that may not reflect the intention of the fog nodes in reality. The security drawbacks of this work are addressed by Zhang et al. by proposing a fog-centric framework that utilizes MAC-based homomorphic tags \cite{zhang}. But this method introduces an additional computational burden for resource-constraint clients. Yoosuf et al. maintain a Bloomier Hash Table (BHT) in the fog layer to perform auditing \cite{yoo}. This approach suffers from an increased probability of hash collision in BHT when the volume of user-generated information grows rapidly. Wang et al. proposed an edge computing-based auditing scheme by maintaining self-balanced binary trees that demands lightweight computation from users \cite{wang}. But the edge devices can be susceptible to man-in-the-middle attacks which can result in eavesdropping on sensitive user data as there’s a lack of an established trust model. 

Because of the improved security, traceability, and immutability, blockchain-based solutions are becoming widely adopted for cloud auditing environments \cite{shaf}\cite{jiax}\cite{ped}\cite{dan}. Multiple solutions are explored in the literature to reduce the redundancy issues that arise with achieving immutability in blockchain-based solutions, which include storing information in an encrypted format, maintaining only hashes on the blockchain while data is stored off-chain, role-based data access control in permissioned blockchains, and so on \cite{fran}. A consortium blockchain solution based on zero-knowledge proofs is proposed in the literature to improve auditability \cite{xu}. Although this system needs to deal with the noticeable synchronization and communication overhead of decentralized systems, the organizational setting for information auditing using Hyperledger Fabric can shift power dynamics towards a centralized group of authority.

\section{Architectural Overview}
\subsection{Merkle Hash Tree}
A Merkle Hash Tree is a binary tree based data structure that includes cryptographic collision-resistant hash functions which can facilitate the integrity and consistency verification of stored information \cite{dahl}\cite{mti}. Each leaf node contains a chunk of information along with one or more properties about the information, and every non-leaf node contains a hash value derived recursively from its child nodes’ hash value by concatenating from left to right and, then applying the predefined hash function. The computed hash values for the nodes make it convenient to generate and validate the proofs of existence and modification for the information stored at the leaf nodes. For instance, in figure 1, a Merkle Tree is formed (denoted by the blue nodes) for file blocks B\textsubscript{1}-B\textsubscript{4}, which is represented under version root 0.

\subsection{Sibling Path}
In a Merkle Hash Tree (MHT), for any leaf node, $l\textsubscript{f} \in MHT$, the sibling path of l\textsubscript{f} consists of the list of hash values in the sibling nodes of each node in the path from l\textsubscript{f} to the root of the MHT. For example, in the initial Merkle Tree depicted by the blue nodes in figure 1, the sibling path of the data block, B\textsubscript{1} contains the hash values in C\textsubscript{2} and C\textsubscript{3-4} nodes. If the hash value and sibling path of any leaf node (l\textsubscript{f}) in the Merkle Tree are known, the intermediate hash values can be calculated to regenerate the root node's hash value in order to verify the integrity of the data residing in the leaf node, l\textsubscript{f}.

\subsection{Achieving Immutability through Persistency}
A data structure is defined as persistent if it can retain prior versions as well as the new version when modifications to the structure are performed. Linked data structures, such as trees and graphs, store and maintain information in ephemeral forms, which implies that when information is updated, the old version is permanently lost \cite{dris}. Persistent data structures can be employed to ensure the capabilities of storing, retrieving, and auditing information in centralized cloud storage systems where user data updates are frequent.  


\begin{figure}[!h]
\centering
\includegraphics[width=3.4in, height=2.0in]{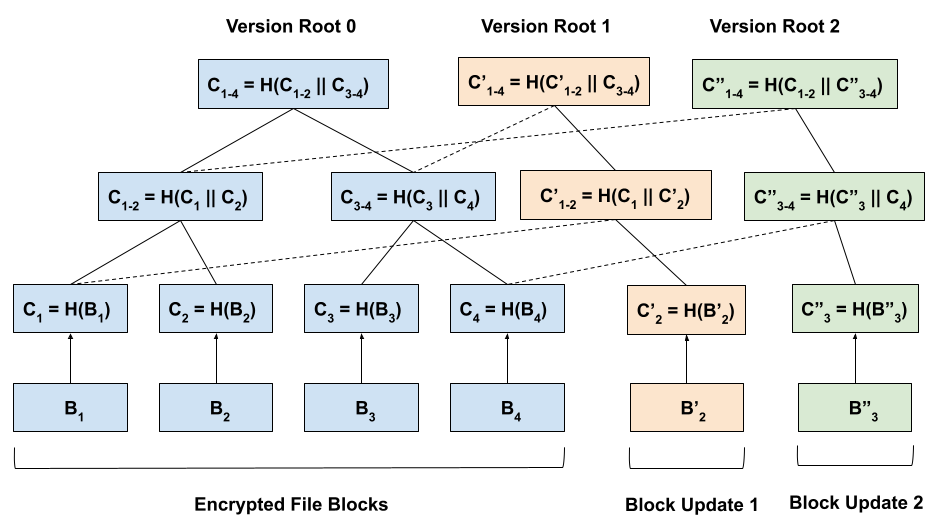}
        \caption{Entangled Merkle Forest Architecture}
\end{figure}

A persistent binary tree-based architecture for cloud service environments has been designed in our proposed paradigm. We leveraged Merkle Tree's features to facilitate information retrieval and auditing through integrity and availability verification, and introduced persistency to Merkle Trees to construct a growing \textbf{Node Sharing Entangled Merkle Forest} in order to achieve immutability. When a file (B\textsubscript{1}-B\textsubscript{4}) is first uploaded to the cloud, an initial Merkle Tree is constructed, which is marked by the blue nodes in figure 1. When block alterations are made to the file, instead of generating whole new trees, only partial trees with selective nodes are constructed. Subsequent updates to B\textsubscript{2} and B\textsubscript{3} blocks with new data blocks B’\textsubscript{2} and B’’\textsubscript{3} results in two new partial trees marked by orange and green nodes, respectively. The trees for different versions share some common nodes to optimize memory and storage costs while retaining and providing access to multiple file versions. In figure 1, versions 0 and 1 of the file share nodes C\textsubscript{1} and C\textsubscript{3-4}, while versions 0 and 2 share nodes C\textsubscript{4} and C\textsubscript{1-2}. When cloud server providers (CSP) and third-party auditors (TPA) are involved, attaining immutability by incorporating persistency in Merkle Trees can provide the capability of version control and dynamic auditing of information in centralized cloud architectures. 
\subsection{System Components}
\textbf{1. Client.} Clients are the interface for users who request services from cloud service providers to store their encrypted files. Before a file is uploaded to the cloud, it is encrypted using a user-generated private key, and then the encrypted file is split into fixed-length blocks at the client. In order to delegate the responsibility of confirming the existence and checking alteration of information in the cloud, the client side informs the Third Party Auditor (TPA) of the file's metadata after initial file uploading and updates TPA every time a file modification is made.

\textbf{2. Third Party Auditor.} Information auditing tasks require intensive computation and communication, which can be a burden for clients with limited resources. Third Party Auditors (TPA) can generate “challenge” messages based on metadata provided by the client in order to validate the existence of files in the cloud on the client's behalf. TPA verifies the correctness of encrypted user data in cloud storage after receiving the “proof” messages from the cloud server. TPA can conduct auditing tasks periodically to evaluate the integrity of different versions of files in the cloud server by challenging the server using client-provided metadata for subsequent file modifications.


\begin{figure}[!h]
\centering
\includegraphics[width=2.0in, height=1.6in]{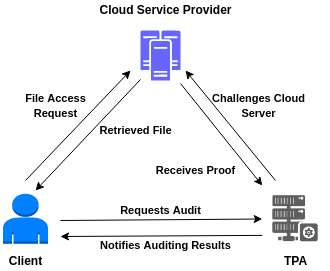}
        \caption{Interactions between System Components}
\end{figure}

\textbf{3. Server.} The primary responsibilities of the server are to store the encrypted files uploaded by clients and to ensure the integrity and consistency of those files. At the server, client-uploaded encrypted file blocks are maintained in the leaf nodes of the series of Merkle trees utilizing our proposed Entangled Merkle Forest architecture. When a file retrieval request is made by the client, the server is able to retrieve the requested version of the file by traversing the relevant tree. 

\section{Implementation Overview}
\subsection{Initial Tree Creation}
A Merkle tree is a hash-based data structure that allows
secure and efficient verification of content in a large file \cite{mti}. To
construct a Merkle tree, we split a file into blocks, group the
blocks into pairs and hash each pair using a collision-resistant
hash function (line 5). The hash values are again grouped in pairs
and each pair is further hashed; this process continues (line 9) until
only a single hash value remains. The last lone hash value is
called the Root of the tree (line 2).\vspace{-0.3 cm} \\

\IncMargin{1em}
\setlength{\textfloatsep}{0pt}
 \begin{algorithm}
  \caption{\textbf{\texttt{CreateInitialTree()}}} 
  \SetKwFunction{BuildTree}{InitialTreeCreation}
  \SetKwInOut{Input}{Input}
  \SetKwInOut{Output}{Output}
  \Indm 
    \Input{List of Encrypted file blocks ($E_b$), Empty tree ($T_b$)} 
    \Output{Each encrypted file block and a calculated hash of the block are stored at a leaf. The root of the created Merkle Tree is returned.}
    \Indp

    \If{\texttt{length($E_b$)} is $1$} {
        \texttt{return} $E_b[0]$\\
    }
        \For{$i=0$ to $\texttt{length(($E_b$)})$ $- 1$}{
            $\texttt{hashPair} \leftarrow \texttt{Hash} (\texttt{$E_b[i]  + E_b[i + 1]$}$) \\
            
            $\texttt{combHashes}.\texttt{push}(\texttt{hashPair})$
            
        $\texttt{$T_b$}.\texttt{push}(\texttt{combHashes})$
      }
    $\texttt{CreateInitialTree}(\texttt{combHashes}, \texttt{$T_b$})$
\end{algorithm}
\DecMargin{1em}


\subsection{Block Update Operation}
By providing the file block number (i\textsubscript{b}), updated block data (b), and the root of the Merkle Tree for the current file state (R\textsubscript{c}) as arguments, the client can update a file block in the cloud server, as shown in algorithm 2. We start building the tree for the modified file state after generating an empty root node (R\textsubscript{u}). In order to achieve immutability, we leverage the current tree state and only create selective nodes that need to be modified in the current state of the tree, rather than constructing a new tree entirely. While traversing the current tree recursively starting from R\textsubscript{c} (lines 4-12, 16-24), if the target leaf node is reached for the block to be updated, a new node is created (lines 2 and 14) instead of modifying the leaf in the current tree and the updated block data is stored in the newly created leaf (line 7 and 19). A fresh node is created for each non-leaf node along the path from the leaf to the root (R\textsubscript{u}). The hash value of every non-leaf node in the updated tree state is calculated from its children and retained (line 27). 

\IncMargin{1em}
\setlength{\textfloatsep}{0pt}
\begin{algorithm}
  \SetKwData{blockID}{\textbf{i\textsubscript{b}}}
  \SetKwData{UpBlock}{\textbf{b}}
  \SetKwData{CurRoot}{\textbf{R\textsubscript{c}}}
  \SetKwData{UpRoot}{\textbf{R\textsubscript{u}}}
  \SetKwData{EmptyNode}{\textbf{n\textsubscript{e}}}
  \SetKwFunction{BuildTree}{InitialTreeCreation}
  \SetKwInOut{Input}{Input}
  \SetKwInOut{Output}{Output}
\Indm  
  \Input{file block index ($i_b$), encrypted block data ($b$) and Merkle tree root for file's current state ($R_c$)}
  \Output{Merkle tree root for file's updated state ($R_u$)}
  \Indp
   
    \eIf {$i_b$ \texttt{is in left half}}{

        $n_e \leftarrow \texttt{new Node}()$\\
        \texttt{Link $n_e$ with the $right$ subtree of the current node}\\
        \For{\texttt{each} $node_i$ $\in$ \texttt{left subtree of} $n_e$}{
            \eIf {$node_i$ \texttt{is a} $leafNode$}{
                $blockHash \leftarrow \texttt{Hash}($$b_i$ in $node_i)$ \\
                $n_e \leftarrow \{b_i, blockHash\}$ \\
            }{
                \texttt{UpdateFileBlock}($i_b$, $b_i$, $R_c$)
            }
        }
    }{
        $n_e \leftarrow \texttt{new Node}()$\\
        \texttt{Link $n_e$ with the $left$ subtree of the current node}\\
        \For{\texttt{each} $node_i$ $\in$ \texttt{right subtree of} $n_e$}{
            \eIf {$node_i$ \texttt{is a} $leafNode$}{
                $blockHash \leftarrow \texttt{Hash}($block $b_i$ in $node_i)$ \\
                $n_e \leftarrow \{b_i, blockHash\}$ \\
            }{
                \texttt{UpdateFileBlock}($i_b$, $b_i$, $R_c$)
            }
        }
    }
    \For{\texttt{each} $node$ $\in$ \texttt{nonLeafNodes}}{
                    $nodeHash \leftarrow \texttt{Hash}(lChild + rChild)$ \\
                }
                $R_u \leftarrow \texttt{CalculateRoot}$()\\
                \texttt{return} $R_u$\\
    \caption{\textbf{\texttt{UpdateFileBlock()}}} 
\end{algorithm}
\DecMargin{1em}

\subsection{Challenge-Proof-Verify}
\textbf{1. Challenge.} The third-party-auditor (TPA) is entrusted with the computation-intensive auditing activities, and the TPA conducts the file auditing responsibilities on behalf of the client and notifies the client of the auditing results. TPA can form ``challenge" messages for the cloud server by selecting a file's version number (v\textsubscript{n}) and a block index (i\textsubscript{b}) at random. To validate the existence of the file states, TPA then sends the challenge messages to the cloud server.

\begin{figure*}
     \centering
     \begin{subfigure}[b]{0.23\textwidth}
         \centering
         \includegraphics[width=1.9in, height=1.6in]{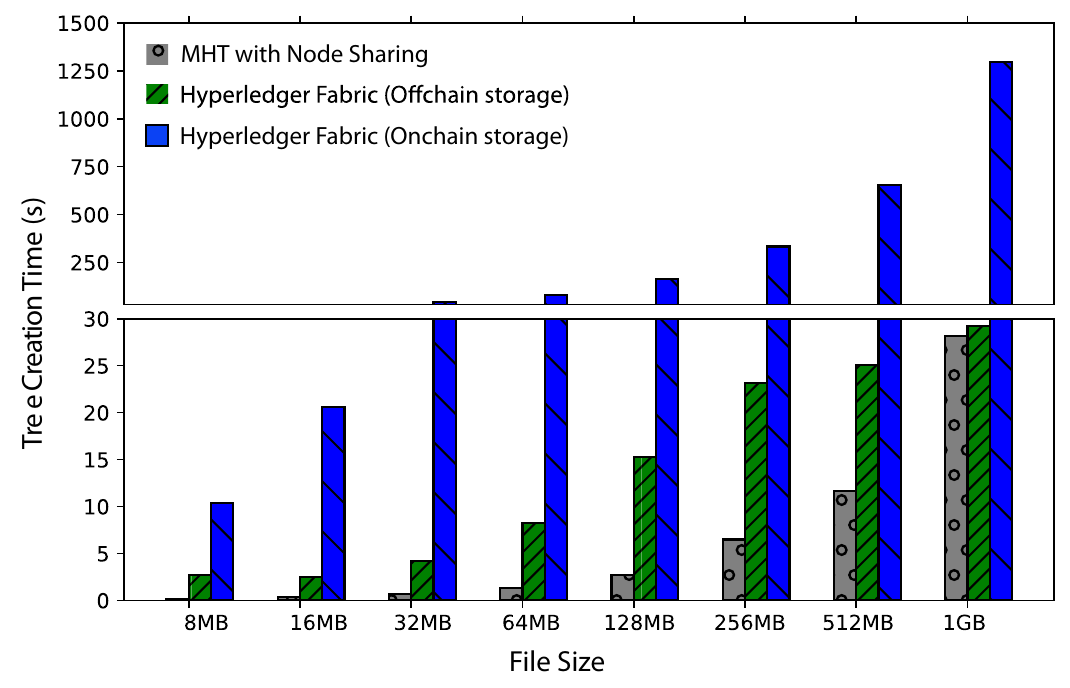}
         \caption{File Sizes vs Initial Tree Creation Tree} \label{fig:performance-a}
     \end{subfigure}
     \hfill
     \begin{subfigure}[b]{0.23\textwidth}
         \centering
         \includegraphics[width=1.9in, height=1.6in]{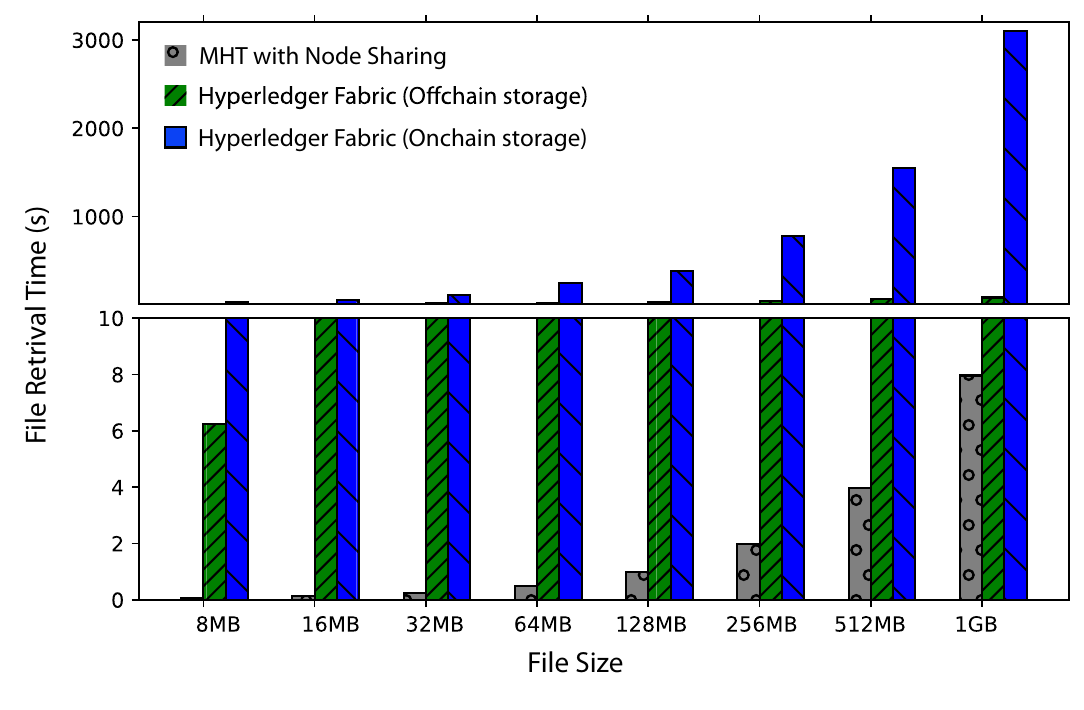}
         \caption{File Sizes vs File Retrieval Time} \label{fig:performance-b}
     \end{subfigure}
     \hfill
     \begin{subfigure}[b]{0.23\textwidth}
         \centering
         \includegraphics[width=1.8in, height=1.6in]{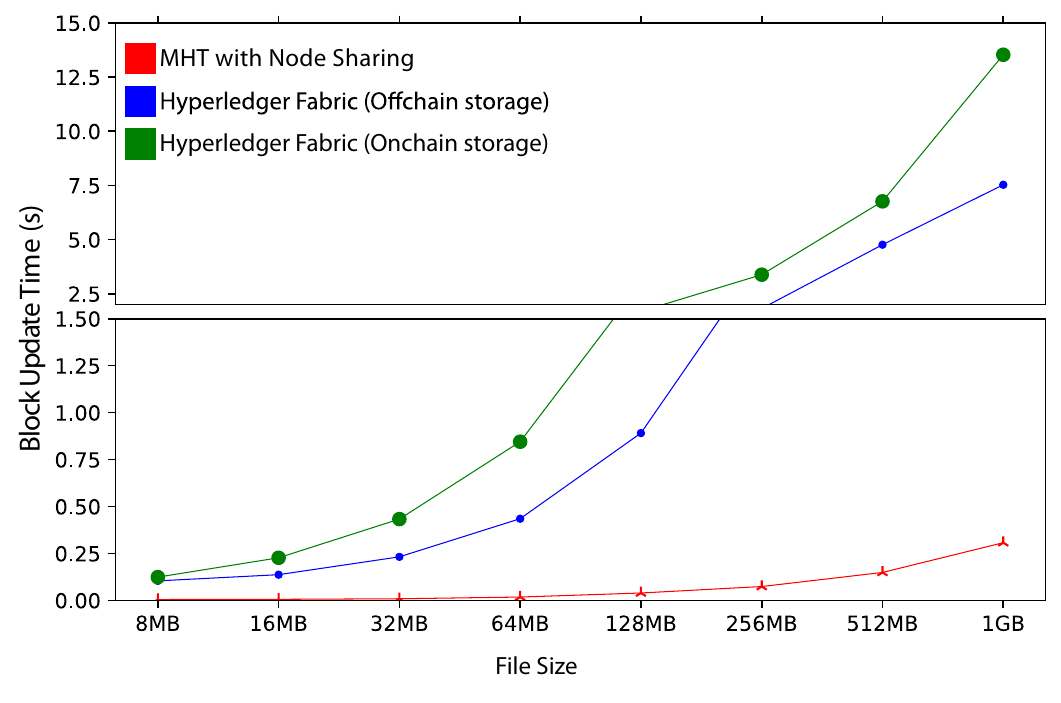}
         \caption{File Sizes vs Block Update Time}
        \label{fig:performance-c}
     \end{subfigure}
     \hfill
     \begin{subfigure}[b]{0.23\textwidth}
         \centering
         \includegraphics[width=1.8in, height=1.6in]{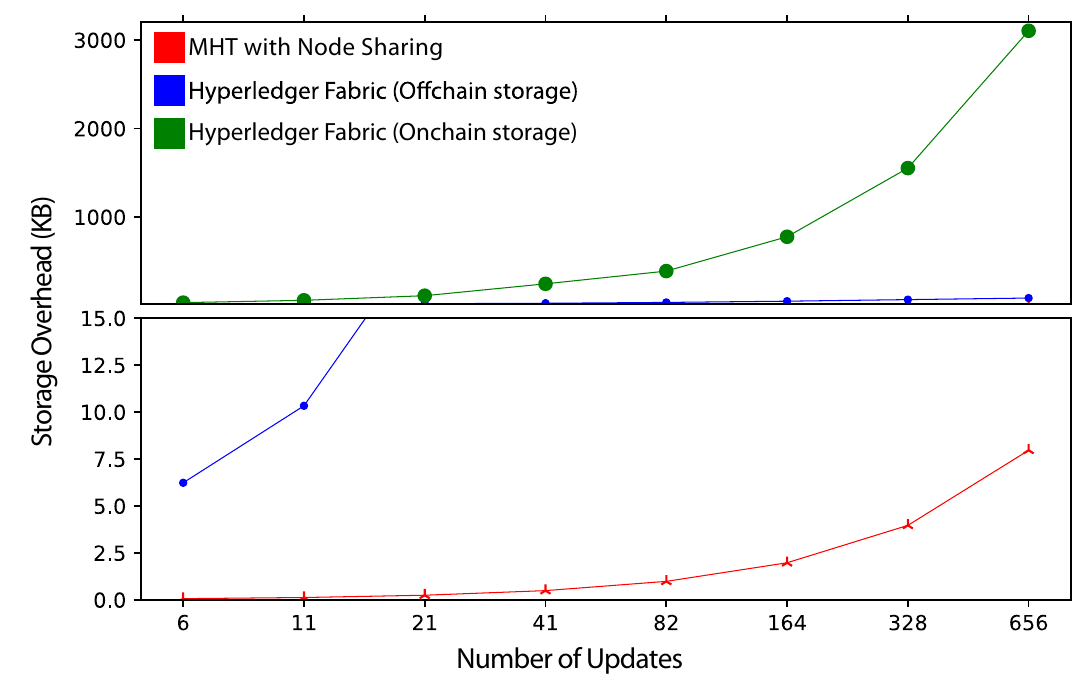}
         \caption{File Version Count vs Storage Overhead} \label{fig:performance-d}
     \end{subfigure}
        \caption{Performance Comparison}
        \label{fig:performance}
\end{figure*}





\textbf{2. Proof.} The server generates a "proof" message using the corresponding root node information when it receives a "challenge" message from the TPA containing a file version number (v\textsubscript{n}) and a block index (i\textsubscript{b}). The stored hashes for all nodes in the sibling path of the block specified by i\textsubscript{b} are included in the proof message. The Merkle Tree for the provided file state (v\textsubscript{n}) is traversed to extract the comprehensive list of hashed values which can be used to regenerate the root node's hash value. For example, when the server is challenged for file state 1 (v\textsubscript{n}=1) and the block {B\textsubscript{2}} in figure 1, the hashed values stored in nodes C\textsubscript{1} and C\textsubscript{3-4} are retrieved and returned to the TPA as a signature of proof to reconstruct C'\textsubscript{1-4} as the version root 1. 

\textbf{3. Verify.} Upon receiving the response at the TPA, the list of hashed values in the sibling path of the block indicated by i\textsubscript{b} is traversed by TPA to construct the root hash value by appending and hashing the values in subsequent indices. The reconstructed root hash is compared to the previously stored client-provided root hash for v\textsubscript{n} at the TPA, and the auditing result is sent to the client. 



    

\subsection{Complexity Analysis} \label{section:complexity}
While conducting the complexity analysis here, the complexity for hashing is omitted. The calculation is based on the number of blocks, $N$ and the number of update operations performed, $Q$. The initial file is divided into $N$ blocks and inserted into the leaf nodes of a Merkle Hash Tree structure. Then $Q$ updates are performed where each update modifies a particular block of file implying to the creation of additional branches in the Entangle Merkle Forest. 

\begin{table}[ht]
\centering

\caption{Space and Time Complexities}
\begin{tabular}{|c|c|c|}
\hline
Operation & Time Complexity & Space Complexity \\
\hline
Initial Tree  & $\mathcal{O}(N)$  & $\mathcal{O}(N)$ \\
 Building & & \\ 
\hline
 Block Update   & $\mathcal{O}(\log_{2}N)$  & $\mathcal{O}(\log_{2}N)$   \\
 Operation & & \\ 
 \hline
 File Retrieval   & $\mathcal{O}(N)$  & $\mathcal{O}(N)$  \\
Operation & & \\ 
 \hline
 Challenge-Proof  & $\mathcal{O}(\log_{2}N+\log_{2}Q)$  & $\mathcal{O}(\log_{2}N)$  \\
 -Verify & & \\ 
 \hline
\end{tabular}

\label{table:complexity}
\end{table}

\section{System Evaluation}
\subsection{Experimental Setup}
We have created an experimental cloud environment to test the prototype of the dynamic auditing framework. Before uploading any file to the cloud, it is split into blocks of fixed 16 KB and then encrypted using Advanced Encryption Standard (AES) scheme at the client end using a user-generated 32 bytes private key. When the file is modified, the third-party auditor (TPA) receives the hash values of the encrypted blocks along with the derived hash of the root node from the client to form challenge messages for the server as metadata. SHA-256 is used as the collision-resistant one-way hash function, making it computationally difficult for curious TPA to obtain information about the original user data. We implemented and evaluated the performance of our scheme in the Ubuntu 22.04 environment using Python3. For a proper comparison with our framework, we choose an ideal case of two baseline blockchain-based auditing schemes-- one with off-chain file storage and the other with on-chain file storage. Blockchain offers immutability as one of its core features. In this evaluation process, we used two Hyperledger Fabric\cite{HyperledgerFabric} based auditing schemes. 



\textbf{Blockchain with Off-chain Storage.} We developed a blockchain-based auditing system using Hyperledger Fabric where a smart contract enables the uploading, downloading, updating, and auditing of cloud files. Similar to the MHT method, this smart contract divides each file into 16KB blocks and applies AES encryption to each block. All the blocks are stored in a separate FTP server, while their hash values are recorded in the blockchain. Whenever an update, insertion, or deletion occurs, a new block with an updated sequence is added to the blockchain.

\textbf{Blockchain with On-chain Storage.} This blockchain network is almost identical to the previous one, with the exception that file blocks are stored directly on the main chain. When updates or deletions take place, the main chain is updated with the new file blocks. Additionally, the file can be retrieved directly from the main chain.


\subsection{Performance Analysis}
We have measured the performance of the Entangled Merkle Forest and compared it with the two blockchain-based schemes considering five metrics: i) initial tree creation time, ii) file retrieval time, iii) block update time, iv) storage overhead, and v) block auditing time.  In Figure \ref{fig:performance-a}, the relationship between the time for initial tree construction and the file size is shown. We used different file sizes ranging from 8MB to 1GB for measuring the required time to create the initial tree. As stated in Section \ref{section:complexity}, the number of blocks increases with the increased file size which eventually shows a linear characteristic. Compared to the blockchain models, our MHT-based approach is way faster. In some cases, the on-chain blockchain model needs more than 40X time to create the initial tree. Similar phenomena can be observed in Figure \ref{fig:performance-b} where the relationship between the file retrieval time and file size is presented. In this case, as well, our suggested method performs better than both on-chain and off-chain blockchain-based models. 

The use of Entangled Merkle Forest with node sharing displays a  significant performance improvement compared to blockchain while performing block update operations, as depicted in Figure \ref{fig:performance-b}. We measured the performance of block updates in batches where each batch contains 1\% of the total blocks in a file. Our proposed model performed 50X faster than the on-chain blockchain and 30X faster than the off-chain blockchain. This is because our approach creates a partial tree instead of an entire tree while updating a file block which justifies the observed patterns of the graphs in Figure \ref{fig:performance-c}. 

The most effective performance improvement is observed in terms of storage overhead for maintaining the auditable files as denoted in Figure \ref{fig:performance-d}. For demonstration, we measured the incremental storage overhead for a 1GB file where subsequent update operations generate new trees to retain the file versions. Our Entangled Merkle Forest architecture requires only a few kilobytes (KB) of overhead for each file update. Compared to the blockchain models this is a great storage optimization for the cloud.

\begin{figure}[t]
\centering
\includegraphics[width=2.9in, height=1.6in]{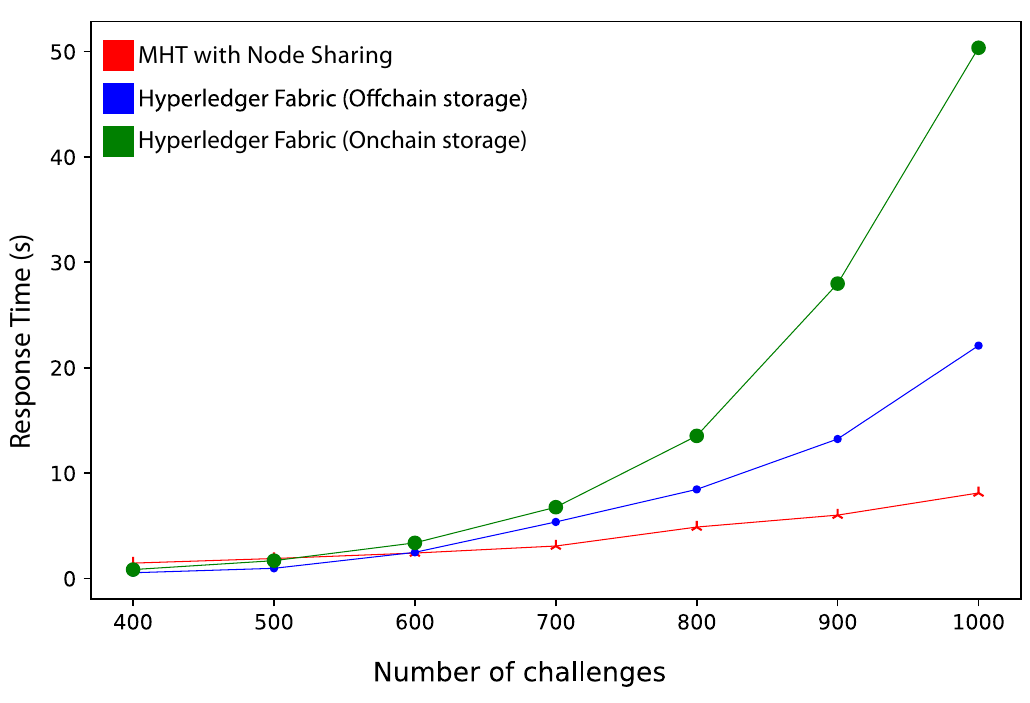}
        \caption{Challenge-Proof-Verify} \label{fig:challenge_response}
\end{figure}
In Figure \ref{fig:challenge_response}, the plot shows the time required (i.e., batch auditing time) for checking the integrity (challenge-proof-verify) of different number of challenged blocks. For a 1GB file, we measured the auditing time ranging from 400 to 1000 blocks. It is evident that, for 400 to 600 blocks, all three schemes take similar auditing time. However, starting from 700 blocks, as the number of blocks increases, both the blockchain-based schemes' performances degrade significantly compared to our scheme. For 1000 blocks, our scheme takes approximately 8 seconds to complete auditing which is ~3X faster (22 seconds) than the off-chain Blockchain and ~6X faster than the on-chain scheme. 

In our scheme, the challenge cases were randomly generated, then the server was challenged. The responses from the server containing the proof messages were recorded and validated, and finally, the average execution times were calculated to create the plot. On the other hand, within the off-chain blockchain network, block verification requires the server to obtain the most recent block number from the main chain, and then retrieve all the file blocks from the challenged block to the latest block from the FTP server, and perform hash recalculation. In the case of the on-chain blockchain, the encrypted file chunks are stored as blocks in the main chain. Similar to off-chain scheme, the on-chain scheme also needs to retrieve all blocks from the challenged block to the latest block while verifying any block integrity. Due to their linear structure, both Blockchain-based schemes take more time than our proposed MHT-based scheme when the number of challenged blocks become larger.

\section{Conclusion}
In this paper, we propose a novel data structure named Entangled Merkle Forest for centralized cloud environments in order to achieve the immutable feature of blockchain, mitigating the synchronization and performance challenges of the decentralized architectures. Immutability is achieved by constructing new tree nodes instead of overwriting the existing ones to form node sharing partial Merkle Trees to retain frequently modified file versions. This proposed paradigm can offer promising research potential for designing secure cloud infrastructures as it can ensure the integrity, privacy, and auditability of cloud data while assuring time and storage efficiency, as demonstrated by our experimental results. \\

\ifCLASSOPTIONcaptionsoff
  \newpage
\fi

\printbibliography

@article{dris,
  title={Making data structures persistent},
  author={Driscoll, James R and Sarnak, Neil and Sleator, Daniel D and Tarjan, Robert E},
  journal={Journal of Computer and System Sciences},
  volume={38},
  pages={86--124},
  year={1989},
  %publisher={Elsevier}
}

@article{wang,
  title={Edge-based auditing method for data security in resource-constrained internet of things},
  author={Wang, Tian and Mei, Yaxin and Liu, Xuxun and Wang, Jin and Dai, HongNing and Wang, Zhijian},
  journal="Journal of Systems Architecture", 
  volume={114},
  pages={101971},
  year={2021},
  publisher={Elsevier}
}

@article{garg,
  title={RITS-MHT: relative indexed and time stamped Merkle hash tree based data auditing protocol for cloud computing},
  author={Garg, Neenu and Bawa, Seema},
  journal={Journal of Network and Computer Applications},
  volume={84},
  pages={1--13},
  year={2017},
  publisher={Elsevier}
}

@article{yu,
  title={An authorized public auditing scheme for dynamic big data storage in cloud computing},
  author={Yu, Han and Lu, Xiuqing and Pan, Zhenkuan},
  journal={IEEE Access},
  volume={8},
  pages={151465--151473},
  year={2020},
  publisher={IEEE}
}

@article{zhang,
  title={Efficient auditing scheme for secure data storage in fog-to-cloud computing},
  author={Zhang, Xingjun and Si, Wei},
  journal={IEEE Access},
  volume={9},
  pages={37951--37960},
  year={2020},
  publisher={IEEE}
}

@article{zhu,
  title={A secure and efficient data integrity verification scheme for cloud-IoT based on short signature},
  author={Zhu, Hongliang and Yuan, Ying and Chen, Yuling and Zha, Yaxing and Xi, Wanying and Jia, Bin and Xin, Yang},
  journal={IEEE Access},
  volume={7},
  pages={90036--90044},
  year={2019},
  publisher={IEEE}
}

@article{fran,
  title={Immutability and decentralized storage: An analysis of emerging threats},
  author={Casino, Fran and Politou, Eugenia and Alepis, Efthimios and Patsakis, Constantinos},
  journal={IEEE Access},
  volume={8},
  pages={4737--4744},
  year={2019},
  publisher={IEEE}
}

@article{jiax,
  title={Blockchain-based public auditing for big data in cloud storage},
  author={Li, Jiaxing and Wu, Jigang and Jiang, Guiyuan and Srikanthan, Thambipillai},
  journal={Information Processing \& Management},
  volume={57},
  number={6},
  pages={102382},
  year={2020},
  publisher={Elsevier}
}

@inproceedings{ped,
  title={Blockchain technology in the auditing environment},
  author={Abreu, Pedro W and Aparicio, Manuela and Costa, Carlos J},
  booktitle={2018 13th Iberian Conference on Information Systems and Technologies (CISTI)},
  pages={1--6},
  year={2018},
  organization={IEEE}
}

@inproceedings{dahl,
  title={Efficient sparse merkle trees},
  author={Dahlberg, Rasmus and Pulls, Tobias and Peeters, Roel},
  booktitle={Nordic Conference on Secure IT Systems},
  pages={199--215},
  year={2016},
  organization={Springer}
}

@inproceedings{mti,
  title={SecReS: A Secure and Reliable Storage Scheme for Cloud with Client-Side Data Deduplication},
  author={Islam, Tariqul and Mistareehi, Hassan and Manivannan, D},
  booktitle={2019 IEEE Global Communications Conference (GLOBECOM)},
  pages={1--6},
  year={2019},
  organization={IEEE}
}

@article{xu,
  title={zkrpChain: Towards multi-party privacy-preserving data auditing for consortium blockchains based on zero-knowledge range proofs},
  author={Xu, Shiwei and Cai, Xiaowen and Zhao, Yizhi and Ren, Zhengwei and Du, Le and Wang, Qin and Zhou, Jianying},
  journal={Future Generation Computer Systems},
  volume={128},
  pages={490--504},
  year={2022},
  publisher={Elsevier}
}

@inproceedings{shaf,
  title={Towards blockchain-based auditable storage and sharing of IoT data},
  author={Shafagh, Hossein and Burkhalter, Lukas and Hithnawi, Anwar and Duquennoy, Simon},
  booktitle={Proceedings of the 2017 on cloud computing security workshop},
  pages={45--50},
  year={2017}
}

@article{deep,
  title={Data provenance in the cloud: A blockchain-based approach},
  author={Tosh, Deepak and Shetty, Sachin and Liang, Xueping and Kamhoua, Charles and Njilla, Laurent L},
  journal={IEEE consumer electronics magazine},
  volume={8},
  number={4},
  pages={38--44},
  year={2019},
  publisher={IEEE}
}

@inproceedings{dan,
  title={Audita: A blockchain-based auditing framework for off-chain storage},
  author={Francati, Danilo and Ateniese, Giuseppe and Faye, Abdoulaye and Milazzo, Andrea Maria and Perillo, Angelo Massimo and Schiatti, Luca and Giordano, Giuseppe},
  booktitle={Proceedings of the Ninth International Workshop on Security in Blockchain and Cloud Computing},
  pages={5--10},
  year={2021}
}

@article{tian,
  title={An efficient and secure data auditing scheme based on fog-to-cloud computing for Internet of things scenarios},
  author={Tian, Jun-Feng and Wang, Hao-Ning},
  journal={International Journal of Distributed Sensor Networks},
  volume={16},
  number={5},
  pages={1550147720916623},
  year={2020},
  publisher={SAGE Publications Sage UK: London, England}
}

@article{yoo,
  title={Lightweight fog-centric auditing scheme to verify integrity of IoT healthcare data in the cloud environment},
  author={Yoosuf, Mohamed Sirajudeen},
  journal={Concurrency and Computation: Practice and Experience},
  volume={33},
  number={24},
  pages={e6450},
  year={2021},
  publisher={Wiley Online Library}
}

@misc{HyperledgerFabric,
  title = {{Hyperledger Fabric} },
  howpublished = {\url{https://www.hyperledger.org/use/fabric}}
}

\AtNextBibliography{\small}
\end{document}